\documentclass[aps,prc,reprint,superscriptaddress,amsmath,amssymb,floatfix]{revtex4-2}

\usepackage{graphicx}
\usepackage{bm}
\usepackage{dcolumn}
\usepackage[colorlinks=true,linkcolor=black,citecolor=black,urlcolor=blue]{hyperref}

\newcommand{\nuc}[2]{$^{#1}$#2}
\newcommand{\Ecm}{E_{\rm c.m.}}
\newcommand{\Ecn}{E^{*}_{\rm CN}}
\newcommand{\Bfus}{B^{*}_{\rm fus}}
\newcommand{\Bqf}{B_{\rm qf}}
\newcommand{\Pcn}{P_{\rm CN}}

\begin{document}

\title{Contribution of incomplete fusion to the evaporation residue cross sections\\ in the \nuc{30}{Si}+\nuc{186}{W} reaction}

\author{Bakhodir Kayumov}
\email{b.kayumov@newuu.uz}
\affiliation{New Uzbekistan University, Tashkent 100000, Uzbekistan}
\affiliation{Institute of Nuclear Physics, Tashkent 100214, Uzbekistan}
\author{Avazbek Nasirov}
\affiliation{Institute of Nuclear Physics, Tashkent 100214, Uzbekistan}
\affiliation{Bogoliubov Laboratory of Theoretical Physics, Joint Institute for Nuclear Research, Dubna 141980, Russia}

\date{\today}

\begin{abstract}
The formation of the evaporation residues in the \nuc{30}{Si}+\nuc{186}{W} reaction leading to the compound nucleus \nuc{216}{Ra}$^*$ has been studied in the framework of the dinuclear system (DNS) model. The capture, complete fusion, quasifission, and incomplete fusion cross sections have been calculated at the collision energies $\Ecm=107.6$--172.2~MeV by averaging over the orientation angles of the axial symmetry axes of the deformed \nuc{30}{Si} and \nuc{186}{W} nuclei. The incomplete fusion is considered as the breakup of the DNS with the emission of the $\alpha$ particle from its very mass-asymmetric configuration, which is populated by the nucleon transfer in collisions with large orbital angular momentum. The survival probabilities of the excited \nuc{216}{Ra} and of the conjugate nucleus \nuc{212}{Rn} formed after the $\alpha$-particle emission have been calculated using the statistical model. It is shown that the complete fusion is the main source of the evaporation residues at the lower collision energies, while above $\Ecm\approx128$~MeV the Rn isotopes formed in the incomplete fusion give more than half of the calculated evaporation residue cross section. The sum of both contributions describes the slow decrease of the measured total evaporation residue cross section at these energies, where the yield of the Ra isotopes decreases by an order of magnitude. The calculation of the de-excitation of \nuc{216}{Ra}$^*$ with the emission of protons and $\alpha$ particles shows that the contribution of the complete fusion to the cross sections of the isotopes \nuc{212}{Rn} and \nuc{211}{Rn} is negligible, and these isotopes are formed in the incomplete fusion. The fusion probability calculated for the partial waves $\ell\leq30\hbar$ is about 0.75 in comparison with the value $0.57\pm0.08$ extracted from the experimental data at the same collision energies.
\end{abstract}

\maketitle

\section{INTRODUCTION}

The formation of a compound nucleus (CN) in heavy-ion collisions is a complex process competing with the quasifission~\cite{Hinde2021,Nasirov2026NPA}. The dinuclear system (DNS)~\cite{Volkov1978,Adamian1998} formed at the capture of the projectile by the target nucleus evolves to the CN by the nucleon transfer between its fragments, but it can break up into two fragments before reaching the equilibrated CN state. This process, called quasifission, decreases the number of events leading to complete fusion, and it is one of the main reasons for the small cross sections of the synthesis of superheavy elements~\cite{Giardina2000,Nasirov2024PRC110,Kayumov2025}. It was believed for a long time that the hindrance to complete fusion becomes significant only in reactions with the product of the charge numbers of the colliding nuclei $Z_1Z_2>1600$~\cite{Blocki1986,Berriman2001,Hinde2021}.

Berriman \textit{et al.}~\cite{Berriman2001} studied the formation of the CN \nuc{216}{Ra}$^*$ in the \nuc{12}{C}+\nuc{204}{Pb}, \nuc{19}{F}+\nuc{197}{Au}, and \nuc{30}{Si}+\nuc{186}{W} reactions with $Z_1Z_2=492$, 711, and 1036, respectively. The beam energies were chosen to form \nuc{216}{Ra}$^*$ at the same excitation energies in the three reactions, and the evaporation residues (ER) were identified by their $\alpha$ decay. The comparison of the cross sections of the Ra isotopes showed that the ER yields in the \nuc{19}{F}+\nuc{197}{Au} and \nuc{30}{Si}+\nuc{186}{W} reactions are $0.64\pm0.09$ and $0.57\pm0.08$ of the one in the \nuc{12}{C}+\nuc{204}{Pb} reaction, respectively~\cite{Berriman2001}. Together with the increase of the width of the mass distribution of the fission fragments with the projectile mass, this result was interpreted as the presence of quasifission in reactions with $Z_1Z_2$ well below 1600~\cite{Berriman2001,Hinde2021}. These conclusions were confirmed by the measurements performed at JINR for the reactions forming \nuc{216}{Ra}$^*$ and other Ra isotopes~\cite{Chizhov2003,Sagaidak2003}. In the liquid-drop picture, the entrance channel of the \nuc{30}{Si}+\nuc{186}{W} reaction lies on the opposite side of the Businaro--Gallone point in comparison with the \nuc{12}{C}+\nuc{204}{Pb} reaction, therefore the DNS formed in this reaction is driven toward the mass-symmetric configurations~\cite{Davies1985,Berriman2001}. In our recent work~\cite{Nasirov2026NPA}, the fusion probability calculated by the DNS model for the very mass-asymmetric \nuc{12}{C}+\nuc{204}{Pb} reaction was found to be close to 1, while it is very small for the less asymmetric \nuc{48}{Ca}+\nuc{168}{Er} reaction leading to the same CN. The influence of the entrance channel on the ER formation was also demonstrated in Ref.~\cite{Sariyal2024}, where the ER cross sections measured in the \nuc{32}{S}+\nuc{154}{Sm} reaction were found to be much larger than the ones in the less mass-asymmetric \nuc{48}{Ti}+\nuc{138}{Ba} reaction leading to the same CN \nuc{186}{Pt}.

It should be noted that the measured cross sections of all ER in the \nuc{30}{Si}+\nuc{186}{W} reaction, which include the Rn and Fr isotopes formed in the $\alpha xn$ and $pxn$ channels, are considerably larger than the ones of the Ra isotopes. Berriman \textit{et al.} pointed out that the yield of all ER could be affected by the breakup of the projectile into $\alpha$-particle clusters leading at large values of the orbital angular momentum $\ell$ to incomplete fusion~\cite{Berriman2001,Vergani1993}, and only the cross sections of the Ra isotopes were used in their analysis. According to the data of Ref.~\cite{Berriman2001}, the contribution of the non-Ra residues to the total ER cross section in the \nuc{30}{Si}+\nuc{186}{W} reaction increases from about 30\% at $\Ecn\approx50$~MeV to about 85\% at $\Ecn\approx83$~MeV. The theoretical explanation of this large yield is the subject of the present work.

The incomplete fusion (ICF) of nuclei in heavy-ion collisions was first observed more than 60 years ago~\cite{Knox1960}. From the analysis of the experimental data it was found that the peripheral collisions are the favorable condition for ICF~\cite{Inamura1977}, and the cross section of the ER formed in ICF increases with the mass asymmetry of the entrance channel~\cite{Morgenstern1984,Kumar2019}. The breakup fusion model~\cite{Udagawa1980} and the sum-rule model~\cite{Wilczynski1980,Wilczynski1982} assume that the projectile breaks up into a light cluster and a conjugate nucleus near the target nucleus, and only the conjugate nucleus fuses with the target. The breakup of the projectile was also studied by combining a classical trajectory model with the stochastic breakup~\cite{VandenBossche2019}. In Refs.~\cite{Nasirov2022PAN,Nasirov2023PLB}, the ICF has been considered as a very mass-asymmetric channel of the quasifission process, and the breakup of the projectile is not assumed. In collisions with large $\ell$, the DNS formed at capture can reach the configuration consisting of the $\alpha$ particle and the heavy conjugate nucleus by the nucleon transfer between its fragments, and this configuration is unstable against breakup. A similar idea of the centrifugal fragmentation was suggested by Volkov~\cite{Volkov2007}. This mechanism allowed us to describe the measured ER cross sections of the ICF channels in the \nuc{18}{O}+\nuc{93}{Nb}~\cite{Agarwal2021,Nasirov2022PAN} and \nuc{22}{Ne}+\nuc{176}{Lu} reactions~\cite{Nasirov2023PLB}, and the inclusion of the ICF channel was necessary to reproduce the ER cross sections measured in the \nuc{32}{S}+\nuc{154}{Sm} reaction at higher energies~\cite{Sariyal2024}.

The main goal of this work is to study the contributions of the complete fusion and ICF to the ER cross sections in the \nuc{30}{Si}+\nuc{186}{W} reaction. The capture, complete fusion, quasifission, and ICF cross sections are calculated by the DNS model~\cite{Nasirov2005,Nasirov2013,Kayumov2022,Nasirov2023PLB}, taking into account the deformation of both colliding nuclei and averaging over their orientation angles. The survival probabilities of the CN \nuc{216}{Ra}$^*$ and of the conjugate nucleus \nuc{212}{Rn}$^*$ formed after the $\alpha$-particle emission are calculated using the statistical model~\cite{Kayumov2025}.

This paper is structured as follows: Sec.~\ref{sec:capture} discusses the capture cross section calculations, Sec.~\ref{sec:fusion} presents the complete fusion and ICF in the DNS model, and Sec.~\ref{sec:er} describes the calculation of the ER cross sections. The results are discussed and compared with the experimental data in Sec.~\ref{sec:results}. Finally, the conclusions are reported in Sec.~\ref{sec:concl}.

\section{CAPTURE CROSS SECTION}\label{sec:capture}

The capture of the projectile by the target nucleus is a necessary condition for both the complete fusion and quasifission. The range of the orbital angular momentum leading to capture is found by solving the dynamical equations of motion for the relative distance $R$ and orbital angular momentum at the given values of $\Ecm$ and orientation angles $\alpha_i$ ($i=1,2$) of the axial symmetry axes of the colliding nuclei relative to the beam direction~\cite{Nasirov2005,Kayumov2022,Nasirov2024PRC110}. Capture occurs when two conditions are satisfied:
\begin{enumerate}
\item[(1)] the collision energy is sufficient to overcome or to tunnel through the entrance-channel barrier of the nucleus-nucleus potential;
\item[(2)] the value of the dissipated kinetic energy corresponds to the size of the potential well, so that the system is trapped in the well~\cite{Kayumov2025}.
\end{enumerate}
The capture cross section is calculated by the formula~\cite{Giardina2000,Kayumov2022}
\begin{equation}
\sigma_{\rm cap}(\Ecm,\ell,\alpha_i)=\frac{\lambda^2}{4\pi}(2\ell+1)P^{(\ell)}_{\rm cap}(\Ecm,\alpha_i),
\label{eq:scap}
\end{equation}
where $\lambda$ is the de Broglie wavelength of the entrance channel. The capture probability is determined by the conditions
\begin{equation}
P^{(\ell)}_{\rm cap}=
\begin{cases}
1, & \ell_m<\ell<\ell_d,\ \Ecm>V_B,\\
0, & \ell<\ell_m \text{ or } \ell>\ell_d,\ \Ecm>V_B,\\
P^{(\ell)}_{\rm tun}, & \text{all } \ell,\ V_{\rm min}<\Ecm\leq V_B,
\end{cases}
\label{eq:pcap}
\end{equation}
where $V_B$ represents the barrier height, while $V_{\rm min}$ denotes the minimum of the potential well of the nucleus-nucleus interaction; $\ell_m$ and $\ell_d$ are the boundary values of the $\ell$ window determined from the solution of the equations of motion. The barrier penetrability $P^{(\ell)}_{\rm tun}$ is calculated by the improved WKB formula of Kemble~\cite{Kemble1935}:
\begin{equation}
P^{(\ell)}_{\rm tun}(\Ecm,\alpha_i)=\frac{1}{1+\exp\left[2K(\Ecm,\ell,\alpha_i)\right]},
\label{eq:ptun}
\end{equation}
\begin{equation}
K(\Ecm,\ell,\alpha_i)=\int_{R_{\rm in}}^{R_{\rm out}}\sqrt{\frac{2\mu}{\hbar^2}\left[V(R,\ell,\alpha_i)-\Ecm\right]}\,dR,
\label{eq:kwkb}
\end{equation}
where $R_{\rm in}$ and $R_{\rm out}$ are the inner and outer turning points determined by the condition $V(R,\ell,\alpha_i)=\Ecm$, and $\mu$ is the reduced mass. The nucleus-nucleus potential $V(R,\ell,\alpha_i)$ is the sum of the Coulomb, nuclear, and rotational parts; its nuclear part is calculated by the double-folding procedure with the effective nucleon-nucleon forces suggested by Migdal~\cite{Migdal1983,Nasirov2005}.

Both colliding nuclei are deformed in their ground states: \nuc{30}{Si} has an oblate shape ($\beta_2=-0.236$) and \nuc{186}{W} has a prolate shape ($\beta_2=0.221$)~\cite{Moller2016}. The deformation parameters of the first $2^+$ and $3^-$ excited states are taken from Refs.~\cite{Raman2001} and \cite{Kibedi2002}, respectively. The values used in the calculations are presented in Table~\ref{tab:def}. The partial cross sections are averaged over the orientation angles of both nuclei:
\begin{eqnarray}
\sigma(\Ecm,\ell)&=&\int_0^{\pi/2}\!\!\int_0^{\pi/2}\sigma(\Ecm,\ell,\alpha_1,\alpha_2)\nonumber\\
&&\times\sin\alpha_1\sin\alpha_2\,d\alpha_1 d\alpha_2,
\label{eq:aver}
\end{eqnarray}
where $\sigma$ stands for the capture, fusion, or ICF partial cross section. In the calculations, seven values $\alpha_i=0^\circ,15^\circ,\ldots,90^\circ$ of each orientation angle have been used.

\begin{table}[b]
\caption{Deformation parameters of the ground state $\beta_2$ and of the first $2^+$ and $3^-$ excited states, and the energies of these states, used in the calculations.}
\label{tab:def}
\begin{ruledtabular}
\begin{tabular}{lccccc}
Nucleus & $\beta_2$~\cite{Moller2016} & $\beta_{2^+}$~\cite{Raman2001} & $E_{2^+}$ (MeV) & $\beta_{3^-}$~\cite{Kibedi2002} & $E_{3^-}$ (MeV)\\
\hline
\nuc{30}{Si}  & $-0.236$ & 0.316 & 2.235 & 0.275 & 5.488\\
\nuc{186}{W}  & $0.221$  & 0.224 & 0.123 & 0.060 & 1.045\\
\end{tabular}
\end{ruledtabular}
\end{table}

\section{COMPLETE FUSION AND INCOMPLETE FUSION}\label{sec:fusion}

\subsection{Potential energy surface}

The DNS formed at capture evolves to complete fusion in competition with the quasifission by the nucleon transfer between its fragments. This evolution is governed by the potential energy surface (PES)~\cite{Nasirov2023PLB,Nasirov2026NPA}:
\begin{equation}
U(Z,A,\ell,R,\{\alpha_i\})=Q_{gg}-V^{\rm CN}_{\rm rot}(\ell)+V(Z,A,\ell,R,\{\alpha_i\}),
\label{eq:pes}
\end{equation}
where $Z$ and $A$ are the charge and mass numbers of a DNS fragment, while the ones of the conjugate fragment are $Z_c=Z_{\rm CN}-Z$ and $A_c=A_{\rm CN}-A$; $Q_{gg}=B_1+B_2-B_{\rm CN}$ is the reaction energy balance, and $B_1$, $B_2$, and $B_{\rm CN}$ are the binding energies of the interacting nuclei and CN obtained from the mass table~\cite{Moller2016}; $V^{\rm CN}_{\rm rot}(\ell)$ is the rotational energy of the CN, and $V(Z,A,\ell,R,\{\alpha_i\})$ is the nucleus-nucleus potential, which includes the Coulomb, nuclear, and rotational energies. The rotational energy of the DNS is calculated with its moment of inertia $J_{\rm DNS}=\mu R_m^2+(J_1+J_2)/2$~\cite{Nasirov2023PLB}:
\begin{equation}
V_{\rm rot}(Z,A,\ell,R)=\frac{\hbar^2\ell(\ell+1)}{2J_{\rm DNS}(Z,A)},
\label{eq:vrot}
\end{equation}
where $\mu$ is the reduced mass of the DNS fragments, $J_1$ and $J_2$ are their moments of inertia, and $R_m$ is the position of the minimum of the potential well. The driving potential $U_{\rm dr}(Z,A,\ell,\{\alpha_i\})$ is the curve connecting the minima of the potential well of the PES along the charge-asymmetry axis.

\begin{figure}[!tb]
\includegraphics[width=\columnwidth]{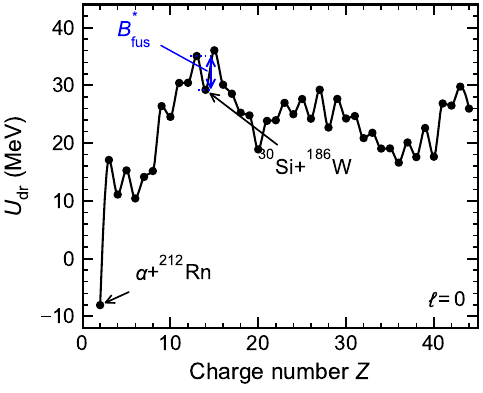}
\caption{Driving potential of the DNS formed in the \nuc{30}{Si}+\nuc{186}{W} reaction calculated for $\ell=0$ as a function of the charge number $Z$ of its light fragment. The intrinsic fusion barrier $\Bfus$ of the entrance channel ($Z_P=14$) and the $\alpha$+\nuc{212}{Rn} configuration ($Z=2$) are indicated.}
\label{fig:udr}
\end{figure}

Figure~\ref{fig:udr} shows the driving potential of the DNS formed in the \nuc{30}{Si}+\nuc{186}{W} reaction calculated for $\ell=0$. The intrinsic fusion barrier $\Bfus$ is the difference between the maximum of the driving potential on the way to complete fusion ($Z\to0$) and its value at the entrance channel $Z_P=14$. It is seen from Fig.~\ref{fig:udr} that $\Bfus=5.86$~MeV. The quasifission barrier $\Bqf$, which is determined by the depth of the potential well of the nucleus-nucleus interaction, is equal to 8.50~MeV for the entrance channel, and the barrier toward the mass-symmetric configurations is $B^*_{\rm sym}=6.84$~MeV. The configuration consisting of the $\alpha$ particle ($Z=2$) and the conjugate nucleus \nuc{212}{Rn} corresponds to the deep minimum of the driving potential with $\Bqf=17.8$~MeV at $\ell=0$. Since the moment of inertia of the very mass-asymmetric DNS is small, its rotational energy (\ref{eq:vrot}) increases strongly with $\ell$ for small values of $Z$. Therefore, in collisions with large $\ell$, the intrinsic fusion barrier increases, while the quasifission barrier of the $\alpha$+\nuc{212}{Rn} configuration decreases~\cite{Nasirov2023PLB}.

\subsection{Evolution of the charge distribution}

The probability $D_Z(E^*_Z,t)$ of the population of the DNS configuration with the charge number $Z$ of its light fragment is found by solving the transport master equation~\cite{Nasirov2016,Nasirov2023PLB}:
\begin{eqnarray}
\frac{\partial D_Z(E^*_Z,t)}{\partial t}&=&\Delta^{(-)}_{Z+1}D_{Z+1}(E^*_Z,t)+\Delta^{(+)}_{Z-1}D_{Z-1}(E^*_Z,t)\nonumber\\
&&-\left(\Delta^{(-)}_{Z}+\Delta^{(+)}_{Z}+\Lambda^{\rm qf}_{Z}\right)D_Z(E^*_Z,t),
\label{eq:master}
\end{eqnarray}
with the initial condition $D_Z(t=0)=1$ for $Z=Z_P$. The transition coefficients $\Delta^{(\pm)}_Z$ of the proton transfer are calculated microscopically, and they depend on the DNS excitation energy $E^*_Z(\ell)$ and on the energies, spins, and occupation numbers of the single-particle states of the nucleons in the DNS fragments~\cite{Nasirov2016,Kayumov2022}. The rate of the DNS breakup through the quasifission barrier $\Bqf(Z,A,\ell,\{\alpha_i\})$ is calculated by the Kramers formula~\cite{Nasirov2013,Nasirov2026NPA}:
\begin{equation}
\Lambda^{\rm qf}_Z=\frac{\Gamma_Z\,\omega_m(Z)}{2\pi\,\omega_{\rm qf}(Z)}\exp\left(-\frac{\Bqf(Z,A,\ell,\{\alpha_i\})}{T_Z}\right),
\label{eq:lambda}
\end{equation}
\begin{equation}
\Gamma_Z=\sqrt{\gamma^2/(2\mu)^2+\omega^2_{\rm qf}(Z)}-\gamma/(2\mu),
\label{eq:gamma}
\end{equation}
where $\omega_m$ and $\omega_{\rm qf}$ are the frequencies of the parabolas used to approximate the potential well and the barrier of the nucleus-nucleus interaction, and $\gamma$ is the friction coefficient of the relative motion. $T_Z=\sqrt{E^*_Z/a}$ is the effective temperature of the DNS with $a=A_{\rm CN}/12$~MeV$^{-1}$, and the DNS excitation energy is
\begin{equation}
E^*_Z(\ell,\{\alpha_i\})=\Ecm-V(Z,A,R_m,\ell,\{\alpha_i\})+\Delta Q_{gg}(Z,A),
\label{eq:exdns}
\end{equation}
where $\Delta Q_{gg}(Z,A)=B_1+B_2-B_P-B_T$ takes into account the change of the intrinsic energy of the DNS due to the nucleon transfer; $B_P$, $B_T$, $B_1$, and $B_2$ are the binding energies of the initial colliding nuclei and of the interacting DNS fragments, respectively.
\begin{figure}[!tb]
\includegraphics[width=\columnwidth]{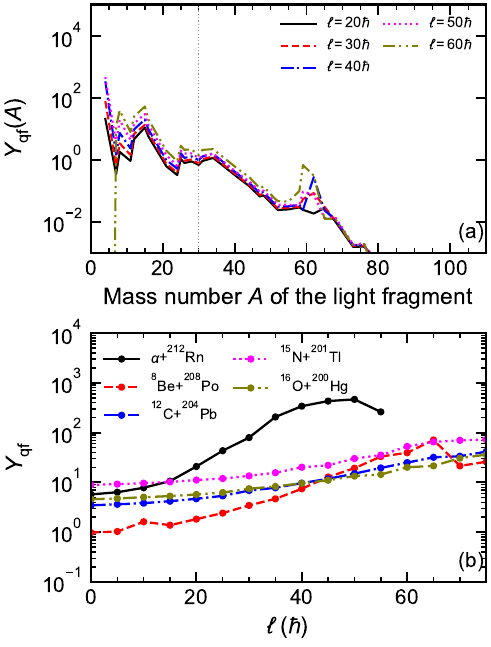}
\caption{(a) Mass distributions of the quasifission products (light fragments) calculated for the \nuc{30}{Si}+\nuc{186}{W} reaction for different values of $\ell$ and averaged over the orientation angles; the dotted line shows the entrance channel. (b) Yields of the quasifission products formed in the breakup of the $\alpha$+\nuc{212}{Rn}, \nuc{8}{Be}+\nuc{208}{Po}, \nuc{12}{C}+\nuc{204}{Pb}, \nuc{15}{N}+\nuc{201}{Tl}, and \nuc{16}{O}+\nuc{200}{Hg} configurations as functions of $\ell$. The yields are given per $10^3$ capture events.}
\label{fig:mass}
\end{figure}
The yield of the quasifission products with the charge number $Z$ is determined by the breakup rate $\Lambda^{\rm qf}_Z$ and the population $D_Z$ of the corresponding DNS configuration [see Eq.~(\ref{eq:yz}) for $Z=2$]. The mass distributions of the quasifission products obtained from the solution of Eq.~(\ref{eq:master}) for the \nuc{30}{Si}+\nuc{186}{W} reaction at different values of $\ell$ and averaged over the orientation angles are presented in Fig.~\ref{fig:mass}(a); the distributions are symmetric with respect to $A_{\rm CN}/2$, and only the light fragments are shown. The DNS formed in the entrance channel ($A=30$) evolves mainly toward the more mass-asymmetric configurations in the direction of the decrease of the driving potential (Fig.~\ref{fig:udr}), and the yield of the mass-symmetric products is small. It should be noted that this result, obtained with the driving potential including the shell effects, differs from the liquid-drop expectation of the evolution of the DNS toward the mass-symmetric configurations in this reaction~\cite{Davies1985,Berriman2001}. The mass distribution has a sharp maximum corresponding to the breakup of the $\alpha$+\nuc{212}{Rn} configuration, which leads to the ICF considered in Sec.~\ref{sec:icf}. The dependence of the yields of the main channels on $\ell$ is shown in Fig.~\ref{fig:mass}(b). The yield of the $\alpha$+\nuc{212}{Rn} channel increases strongly with $\ell$, since the quasifission barrier of this configuration decreases with the increase of its rotational energy, while the yields of the other light products increase more gradually. At $\ell\geq60\hbar$, the excitation energy of the $\alpha$+\nuc{212}{Rn} configuration is lower than its quasifission barrier, the $\alpha$-particle emission does not take place, and the quasifission products are formed mainly in the configurations with the light fragments $A\approx7$--16.

\subsection{Fusion probability}

The fusion probability is determined by the charge distribution $D_Z$ of the DNS fragments~\cite{Kayumov2022,Kayumov2025}:
\begin{equation}
\Pcn(\Ecm,\ell)=\sum_{Z=2}^{Z_{\rm CN}/2}D_Z(E^*_Z,\ell)\,P^{(Z)}_{\rm CN}(E^*_Z,\ell),
\label{eq:pcn}
\end{equation}
where the summation is performed over all charge configurations of the light DNS fragment, and the dependence on the orientation angles $\{\alpha_i\}$ is not indicated for simplicity. The fusion probability $P^{(Z)}_{\rm CN}$ from the charge configuration $Z$ is calculated as the branching ratio of the level densities of the Fermi system above the intrinsic fusion barrier $B^{*(Z)}_{\rm fus}$, the quasifission barrier $B^{(Z)}_{\rm qf}$, and the barrier $B^{*(Z)}_{\rm sym}$ toward the mass-symmetric configurations~\cite{Nasirov2019}:
\begin{equation}
P^{(Z)}_{\rm CN}=\frac{\rho_{\rm fus}}{\rho_{\rm fus}+\rho_{\rm qf}+\rho_{\rm sym}},
\label{eq:pcnz}
\end{equation}
where $\rho_{\rm fus}=\rho(E^*_Z-B^{*(Z)}_{\rm fus})$, $\rho_{\rm qf}=\rho(E^*_Z-B^{(Z)}_{\rm qf})$, and $\rho_{\rm sym}=\rho(E^*_Z-B^{*(Z)}_{\rm sym})$; the level density of the Fermi gas $\rho(E^*)\propto\exp(2\sqrt{aE^*})$ with $a=A_{\rm CN}/10$~MeV$^{-1}$ is used.
The partial cross section of complete fusion is
\begin{eqnarray}
\sigma_{\rm fus}(\Ecm,\ell,\{\alpha_i\})&=&\sigma_{\rm cap}(\Ecm,\ell,\{\alpha_i\})\nonumber\\
&&\times\Pcn(\Ecm,\ell,\{\alpha_i\}),
\label{eq:sfus}
\end{eqnarray}
and it is averaged over the orientation angles by Eq.~(\ref{eq:aver}). The fast-fission cross section is calculated for the partial waves $\ell_B<\ell<\ell_d$~\cite{Nasirov2024PRC110}, where $\ell_B\approx73\hbar$ is the value at which the macroscopic fission barrier of \nuc{216}{Ra} vanishes~\cite{Sierk1986}. For the \nuc{30}{Si}+\nuc{186}{W} reaction, the fast-fission cross section does not exceed 0.3~mb in the considered energy range.

\subsection{Incomplete fusion}\label{sec:icf}

The part of $D_Z$ corresponding to the breakup of the DNS with $Z=2$ is a contribution to the ICF~\cite{Nasirov2022PAN,Nasirov2023PLB}. The probability of the $\alpha$-particle emission is calculated by the formula
\begin{equation}
Y_{2}(\Ecm,\ell)=\sum_{k=0}^{k_{\rm max}}\Lambda^{\rm qf}_{2}\,D_{2}(E^*_2,\ell,k\Delta t)\,\Delta t,
\label{eq:yz}
\end{equation}
where $\Delta t=10^{-22}$~s, and $k_{\rm max}\Delta t$ is the interaction time after which $D_Z<10^{-5}$, i.e., the DNS has gone to complete fusion or it has broken up. The partial ICF cross section is
\begin{equation}
\sigma_{\rm ICF}(\Ecm,\ell)=\sigma_{\rm cap}(\Ecm,\ell)\,Y_{2}(\Ecm,\ell),
\label{eq:sicf}
\end{equation}
and it is averaged over the orientation angles by Eq.~(\ref{eq:aver}). At the DNS breakup the $\alpha$ particle overcomes the quasifission barrier, and the excitation energy of the conjugate nucleus \nuc{212}{Rn} is~\cite{Nasirov2023PLB}
\begin{equation}
E^*_{\rm ICF}=E^*_{Z=2}-\Bqf(Z=2,\ell,\{\alpha_i\}).
\label{eq:eicf}
\end{equation}
The angular momentum $L_c$ of the conjugate nucleus is estimated as in Ref.~\cite{Nasirov2023PLB} from the sharing of the angular momentum between the relative motion and the rotation of the DNS fragments; in the present calculations $L_c\approx(0.20$--$0.25)\ell$.

\section{EVAPORATION RESIDUE CROSS SECTIONS}\label{sec:er}

The partial cross section of the ER formed after the emission of $x$ neutrons is calculated at each step of the de-excitation cascade by the formula~\cite{Mandaglio2012,Kayumov2025}
\begin{equation}
\sigma^{(x)}_{\rm ER}(E^*_x,\ell)=\sigma^{(x-1)}_{\rm ER}(E^*_{x-1},\ell)\,W^{(x)}_{\rm sur}(E^*_{x-1},\ell),
\label{eq:ser}
\end{equation}
where $\sigma^{(x-1)}_{\rm ER}(E^*_{x-1},\ell)$ is the partial cross section of the formation of the intermediate excited nucleus at the $(x-1)$th step, and $W^{(x)}_{\rm sur}$ is its survival probability against fission. For the complete fusion, $\sigma^{(0)}_{\rm ER}=\sigma_{\rm fus}(\Ecn,\ell)$ with $\Ecn=\Ecm+Q$ and $Q=-70.22$~MeV for the \nuc{30}{Si}+\nuc{186}{W} reaction. For the ICF, $\sigma^{(0)}_{\rm ER}=\sigma_{\rm ICF}(E^*_{\rm ICF},L_c)$ of the conjugate nucleus \nuc{212}{Rn}$^*$. The ER cross section of the $xn$ channel is obtained by summing the partial cross sections over $\ell$, and the total ER cross section is
\begin{equation}
\sigma^{\rm tot}_{\rm ER}=\sum_{x}\sigma^{(x)}_{\rm ER}({\rm CF})+\sum_{x}\sigma^{(x)}_{\rm ER}({\rm ICF}),
\label{eq:stot}
\end{equation}
where the first sum gives the Ra isotopes ($xn$ channels of \nuc{216}{Ra}$^*$) and the second one gives the Rn isotopes ($\alpha xn$ channels); the $\alpha0n$ channel corresponds to \nuc{212}{Rn} formed when $0\leq E^*_{\rm ICF}<S_n(^{212}{\rm Rn})$.

The survival probabilities are calculated using the statistical model as described in Ref.~\cite{Kayumov2025}. The neutron emission widths are calculated in the Weisskopf--Ewing approximation~\cite{Weisskopf1937,WeisskopfEwing1940} with the optical potential of Koning and Delaroche~\cite{Koning2003}, and the fission width is calculated by the Bohr--Wheeler transition-state model~\cite{Bohr1939} corrected by the Kramers factor~\cite{Kramers1940} with the reduced friction coefficient $\beta=5\times10^{21}$~s$^{-1}$ and by the Strutinsky factor~\cite{Strutinsky1973}; the curvatures of the potential at the ground state and at the saddle point are equal to 1~MeV. The level density parameter is calculated by the expression of Nerlo-Pomorska \textit{et al.}~\cite{NerloPomorska2006}, and the shell damping is included according to Ignatyuk's prescription~\cite{Ignatyuk1975} with the damping energy $E_d=19$~MeV; the ratio $a_f/a_n=1$ is used. The pairing and collective enhancement of the level density and the $\gamma$-ray emission are taken into account. The fission barriers are taken from the table of M\"oller \textit{et al.}~\cite{Moller2015}: $B_f=13.01$~MeV for \nuc{216}{Ra} and 18.63~MeV for \nuc{212}{Rn}. The masses and shell corrections are taken from Ref.~\cite{Moller2016}. The ER cross sections presented in Table~\ref{tab:res} and Figs.~\ref{fig:ercf} and \ref{fig:ertot} are calculated taking into account the neutron emission; the contribution of the emission of protons and $\alpha$ particles from \nuc{216}{Ra}$^*$ is discussed in Sec.~\ref{sec:erres}.

\section{RESULTS AND DISCUSSION}\label{sec:results}

\subsection{Capture, fusion, and incomplete fusion cross sections}

\begin{figure}[!tb]
\includegraphics[width=\columnwidth]{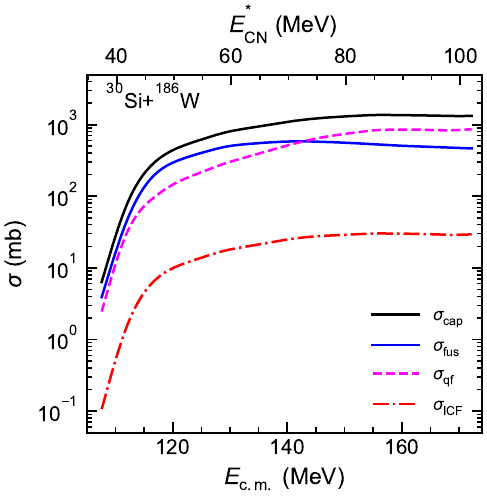}
\caption{Capture ($\sigma_{\rm cap}$), complete fusion ($\sigma_{\rm fus}$), quasifission ($\sigma_{\rm qf}$), and incomplete fusion ($\sigma_{\rm ICF}$) cross sections calculated for the \nuc{30}{Si}+\nuc{186}{W} reaction as functions of $\Ecm$. The upper axis shows the excitation energy of the CN \nuc{216}{Ra}$^*$.}
\label{fig:xs}
\end{figure}

Figure~\ref{fig:xs} presents the capture, complete fusion, quasifission, and ICF cross sections calculated for the \nuc{30}{Si}+\nuc{186}{W} reaction as functions of $\Ecm$. The capture cross section increases with the collision energy and saturates at about 1.3~b. The complete fusion cross section reaches its maximum near $\Ecm\approx142$~MeV and then decreases slowly, while the quasifission cross section continues to increase and becomes larger than $\sigma_{\rm fus}$ at higher energies (Table~\ref{tab:res}). This result is related to the increase of the contribution of the large partial waves to capture: the intrinsic fusion barrier $\Bfus$ increases with $\ell$, and the DNS formed at large $\ell$ breaks up into two fragments before reaching the CN state.

The ICF cross section increases by two orders of magnitude from the lowest collision energy up to $\Ecm\approx120$~MeV and reaches about 30~mb at higher energies (Table~\ref{tab:res}). It should be noted that the ICF cross section is a part of the quasifission cross section, since in the DNS model the ICF is its very mass-asymmetric channel~\cite{Nasirov2022PAN,Nasirov2023PLB}. The ratio $\sigma_{\rm ICF}/\sigma_{\rm cap}$, which can be considered as the probability of the $\alpha$-particle emission from the DNS, is shown in Fig.~\ref{fig:pcn}. It is about 0.02 at $\Ecm\geq116$~MeV and depends weakly on the collision energy. However, the values of $\sigma_{\rm ICF}$ presented in Table~\ref{tab:res} include also the partial waves with $E^*_{\rm ICF}<0$, which do not lead to the ER, as discussed in Sec.~\ref{sec:ldist}.

\subsection{Fusion probability}

\begin{figure}[!tb]
\includegraphics[width=\columnwidth]{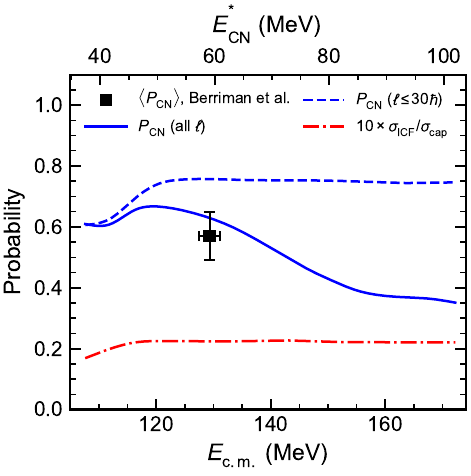}
\caption{Fusion probability $\Pcn$ calculated for the \nuc{30}{Si}+\nuc{186}{W} reaction for all partial waves (solid curve) and for $\ell\leq30\hbar$ (dashed curve), and the ratio $\sigma_{\rm ICF}/\sigma_{\rm cap}$ multiplied by 10 (dash-dotted curve). The square with error bars shows the value $\langle\Pcn\rangle=0.57\pm0.08$ obtained in Ref.~\cite{Berriman2001} from the ratio of the yields of the Ra isotopes in the \nuc{30}{Si}+\nuc{186}{W} and \nuc{12}{C}+\nuc{204}{Pb} reactions at the two highest energies of the \nuc{12}{C}+\nuc{204}{Pb} measurement.}
\label{fig:pcn}
\end{figure}

The fusion probability $\Pcn=\sigma_{\rm fus}/\sigma_{\rm cap}$ calculated for the \nuc{30}{Si}+\nuc{186}{W} reaction is shown in Fig.~\ref{fig:pcn} by the solid curve. It is about 0.6--0.7 at the lower collision energies and decreases to 0.35 at the highest energy due to the increasing contribution of the large partial waves, where the hindrance to complete fusion is strong. The ER are formed mainly in the collisions with small $\ell$, since the fission probability of \nuc{216}{Ra}$^*$ exceeds 98\% already at $\ell=30\hbar$~\cite{Berriman2001}. Therefore, the fusion probability averaged over the partial waves $\ell\leq30\hbar$ (dashed curve in Fig.~\ref{fig:pcn}) should be compared with the value $\langle\Pcn\rangle=0.57\pm0.08$ extracted by Berriman \textit{et al.} from the ratio of the yields of the Ra isotopes in the \nuc{30}{Si}+\nuc{186}{W} and \nuc{12}{C}+\nuc{204}{Pb} reactions~\cite{Berriman2001,Hinde2021}. This value has been obtained at the two highest energies of the \nuc{12}{C}+\nuc{204}{Pb} measurement, $\Ecn\approx57$--61~MeV, which correspond to $\Ecm\approx127$--131~MeV for the \nuc{30}{Si}+\nuc{186}{W} reaction. Since the fusion probability calculated by the DNS model for the \nuc{12}{C}+\nuc{204}{Pb} reaction is close to 1~\cite{Nasirov2026NPA}, this ratio can be considered as the fusion probability of the \nuc{30}{Si}+\nuc{186}{W} reaction. The calculated value of $\Pcn(\ell\leq30\hbar)$ is nearly constant above $\Ecm\approx120$~MeV, and at $\Ecm\approx127$--131~MeV it is about 0.75, i.e., it exceeds the experimental value by about 0.2.

\subsection{Angular momentum distributions}\label{sec:ldist}

\begin{figure}[!tb]
\includegraphics[width=\columnwidth]{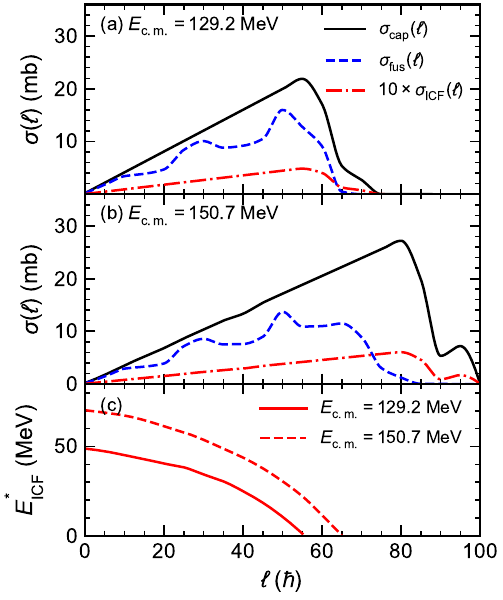}
\caption{Partial capture, complete fusion, and ICF cross sections calculated for the \nuc{30}{Si}+\nuc{186}{W} reaction at (a) $\Ecm=129.2$~MeV and (b) $\Ecm=150.7$~MeV. The bumps of $\sigma_{\rm fus}(\ell)$ at $\ell\approx30\hbar$ and $50\hbar$ are related to the change of the orientation-averaged intrinsic fusion barrier $\Bfus$ with $\ell$. The ICF cross sections are multiplied by 10. In panel (b), the partial waves $\ell\geq70\hbar$, including the maximum of $\sigma_{\rm ICF}(\ell)$ at $\ell\approx80\hbar$, correspond to $E^*_{\rm ICF}<0$, and they do not contribute to the ER. (c) Excitation energy $E^*_{\rm ICF}$ of the conjugate nucleus \nuc{212}{Rn} given by Eq.~(\ref{eq:eicf}) for the partial waves.}
\label{fig:partial}
\end{figure}

The angular momentum distributions of the capture, complete fusion, and ICF cross sections calculated for $\Ecm=129.2$ and 150.7~MeV are presented in Figs.~\ref{fig:partial}(a) and \ref{fig:partial}(b). The partial ICF cross section increases with $\ell$, and the main part of the ICF events leading to the ER is formed at $\ell=20$--$60\hbar$, i.e., in the range of large orbital angular momentum, in agreement with the conclusions of Refs.~\cite{Inamura1977,Wilczynski1980,Nasirov2022PAN,Nasirov2023PLB}. In this range of $\ell$, the hindrance of the complete fusion by the quasifission increases with $\ell$. The excitation energy $E^*_{\rm ICF}$ of the conjugate nucleus \nuc{212}{Rn} decreases with $\ell$ [Fig.~\ref{fig:partial}(c)], since the rotational energy of the DNS increases and its excitation energy $E^*_Z$ decreases. Consequently, the conjugate nuclei formed in the ICF are much colder than \nuc{216}{Ra}$^*$ formed at the same collision energy (Table~\ref{tab:res}): the mean value of $E^*_{\rm ICF}$ increases from 23~MeV at $\Ecm=120.6$~MeV to 33~MeV at 150.7~MeV, and the mean angular momentum of \nuc{212}{Rn}$^*$ is $L_c\approx7$--$9\hbar$.

At $\Ecm=150.7$~MeV, the maximum of $\sigma_{\rm ICF}(\ell)$ is located at $\ell\approx80\hbar$, near the upper boundary of the capture window. For the partial waves $\ell\geq70\hbar$ at this energy, as well as for $\ell\geq60\hbar$ at $\Ecm=129.2$~MeV, the DNS excitation energy is lower than the quasifission barrier of the $\alpha$+\nuc{212}{Rn} configuration, and Eq.~(\ref{eq:eicf}) gives negative values. The survival probability of the conjugate nucleus is equal to zero for these events, therefore they do not contribute to the ER cross sections, although they give up to about half of $\sigma_{\rm ICF}$ at the highest energies. Without these events, the ICF cross section is about 0.02 of $\sigma_{\rm cap}$ at $\Ecm=116$--130~MeV, and it decreases to 0.011--0.016 of $\sigma_{\rm cap}$ at higher energies.

\subsection{Evaporation residue cross sections}\label{sec:erres}

\begin{table*}[t]
\caption{Calculated cross sections (in mb) of capture, complete fusion, quasifission, and ICF, the fusion probability $\Pcn=\sigma_{\rm fus}/\sigma_{\rm cap}$, and the ER cross sections formed in the complete fusion (Ra isotopes) and ICF (Rn isotopes) for the \nuc{30}{Si}+\nuc{186}{W} reaction. $\Ecn=\Ecm+Q$ with $Q=-70.22$~MeV.}
\label{tab:res}
\begin{ruledtabular}
\begin{tabular}{ccccccccccc}
$E_{\rm lab}$ (MeV) & $\Ecm$ (MeV) & $\Ecn$ (MeV) & $\sigma_{\rm cap}$ & $\sigma_{\rm fus}$ & $\sigma_{\rm qf}$ & $\Pcn$ & $\sigma_{\rm ICF}$ & $\sigma_{\rm ER}({\rm CF})$ & $\sigma_{\rm ER}({\rm ICF})$ & $\sigma^{\rm tot}_{\rm ER}$\\
\hline
125 & 107.6 & 37.4 & 5.98 & 3.71 & 2.27 & 0.62 & 0.100 & 1.27 & 0.087 & 1.36 \\
130 & 111.9 & 41.7 & 80.8 & 46.6 & 34.2 & 0.58 & 1.61 & 8.66 & 1.44 & 10.1 \\
135 & 116.2 & 46.0 & 262 & 178 & 83.4 & 0.68 & 5.85 & 20.3 & 4.97 & 25.3 \\
140 & 120.6 & 50.3 & 459 & 303 & 156 & 0.66 & 10.3 & 17.1 & 8.07 & 25.2 \\
145 & 124.9 & 54.6 & 607 & 396 & 212 & 0.65 & 13.5 & 12.6 & 9.02 & 21.6 \\
150 & 129.2 & 59.0 & 789 & 497 & 292 & 0.63 & 17.8 & 8.22 & 10.2 & 18.4 \\
155 & 133.5 & 63.2 & 896 & 537 & 359 & 0.60 & 19.9 & 4.75 & 8.74 & 13.5 \\
160 & 137.8 & 67.6 & 1030 & 570 & 460 & 0.55 & 23.3 & 3.02 & 7.30 & 10.3 \\
165 & 142.1 & 71.9 & 1153 & 588 & 566 & 0.51 & 26.3 & 2.03 & 8.54 & 10.6 \\
170 & 146.4 & 76.2 & 1248 & 572 & 676 & 0.46 & 28.0 & 1.32 & 7.26 & 8.58 \\
175 & 150.7 & 80.5 & 1306 & 558 & 748 & 0.43 & 29.0 & 0.984 & 8.44 & 9.42 \\
180 & 155.0 & 84.8 & 1371 & 530 & 840 & 0.39 & 30.3 & 0.805 & 7.33 & 8.14 \\
185 & 159.3 & 89.1 & 1354 & 508 & 846 & 0.38 & 30.0 & 0.765 & 6.16 & 6.92 \\
190 & 163.6 & 93.4 & 1352 & 496 & 855 & 0.37 & 30.0 & 0.816 & 5.20 & 6.02 \\
195 & 167.9 & 97.7 & 1302 & 478 & 823 & 0.37 & 28.7 & 0.951 & 9.41 & 10.4 \\
200 & 172.2 & 102.0 & 1332 & 466 & 866 & 0.35 & 29.4 & 1.16 & 11.1 & 12.3 \\
\end{tabular}
\end{ruledtabular}
\end{table*}

\begin{figure}[!tb]
\includegraphics[width=\columnwidth]{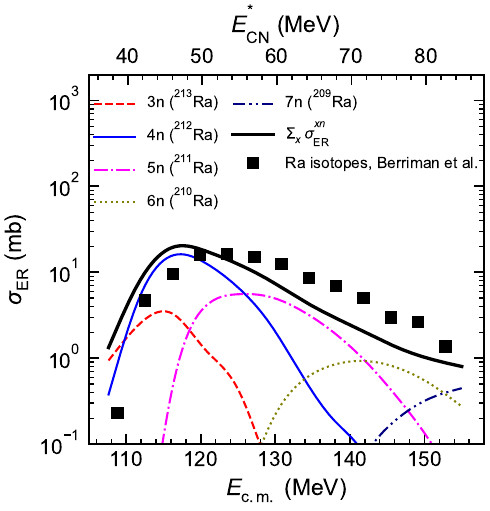}
\caption{ER cross sections of the $xn$ channels calculated for the complete fusion in the \nuc{30}{Si}+\nuc{186}{W} reaction (thin curves) and their sum over $x=1$--10 (thick curve) in comparison with the cross sections of the Ra isotopes measured by Berriman \textit{et al.}~\cite{Berriman2001} (squares).}
\label{fig:ercf}
\end{figure}

The capture, complete fusion, quasifission, and ICF cross sections, the fusion probability, and the ER cross sections formed in the complete fusion and ICF calculated for the \nuc{30}{Si}+\nuc{186}{W} reaction are listed in Table~\ref{tab:res}. Figure~\ref{fig:ercf} presents the ER cross sections of the $xn$ channels calculated for the complete fusion in comparison with the cross sections of the Ra isotopes measured by Berriman \textit{et al.}~\cite{Berriman2001}. The experimental values have been derived from the reduced cross sections presented in Fig.~2 of Ref.~\cite{Berriman2001}, and the collision energy has been obtained as $\Ecm=\Ecn-Q$. The 4$n$ channel dominates near the maximum of the excitation function, while above $\Ecm\approx130$~MeV the main contribution is given by the 5$n$ channel and, at higher energies, by the 6$n$ and 7$n$ channels. The sum of the $xn$ cross sections reproduces the magnitude of the measured cross sections of the Ra isotopes: the calculated maximum is 20.3~mb at $\Ecm=116.2$~MeV in comparison with the measured maximum of 16.3~mb at about 123.5~MeV. The calculation overestimates the data on the low-energy side of the maximum and underestimates them by a factor of 1.5--2.5 on the high-energy side. It should be noted that the position and height of the maximum of the calculated excitation function are sensitive to the fission barriers and shell corrections, which are taken from the theoretical tables~\cite{Moller2015,Moller2016}, and to the level-density parameters used in the statistical model. The strong dependence of the survival probability on the mass model used in the calculations has been demonstrated recently in Ref.~\cite{Kawai2026}. In addition, the experimental excitation energies have an uncertainty related to the beam energy and its loss in the target.

\begin{figure}[!tb]
\includegraphics[width=\columnwidth]{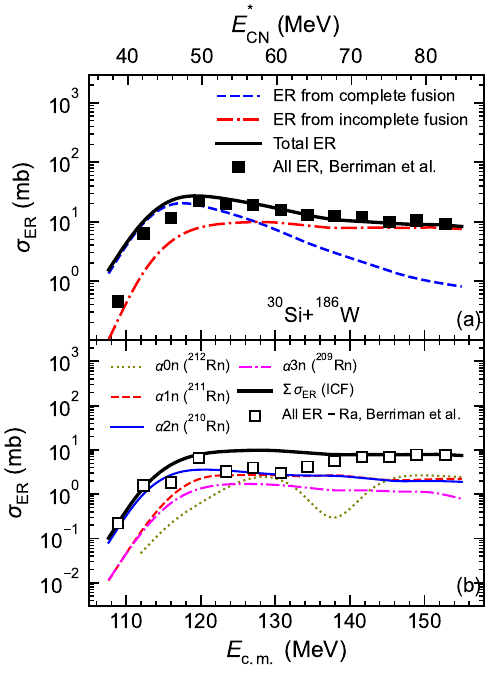}
\caption{(a) ER cross sections calculated for the complete fusion (dashed curve) and ICF (dash-dotted curve) in the \nuc{30}{Si}+\nuc{186}{W} reaction, and their sum (solid curve), in comparison with the measured cross sections of all ER~\cite{Berriman2001} (squares). (b) ER cross sections of the $\alpha xn$ channels ($x=0$--3) formed in the ICF (thin curves) and their sum over $x=0$--10 (thick curve) in comparison with the difference between the measured cross sections of all ER and of the Ra isotopes~\cite{Berriman2001} (open squares). The minimum of the $\alpha0n$ cross section near $\Ecm=138$~MeV is related to the step $\Delta\ell=5\hbar$ of the partial waves used in the calculation.}
\label{fig:ertot}
\end{figure}

Figure~\ref{fig:ertot}(a) compares the calculated total ER cross section (\ref{eq:stot}) with the measured cross section of all ER~\cite{Berriman2001}. The contribution of the complete fusion dominates at $\Ecm\leq125$~MeV, while the ER formed in the ICF give more than half of the calculated ER cross section above $\Ecm\approx128$~MeV. Owing to this contribution, the calculated total ER cross section decreases slowly at higher energies, as it is observed in the experiment, while the cross section of the Ra isotopes decreases by an order of magnitude in this energy range (Fig.~\ref{fig:ercf}). The calculated total cross section agrees with the measured data within 17\% at $\Ecm=120$--151~MeV, and it overestimates the data at $\Ecm\leq116$~MeV in the same way as in Fig.~\ref{fig:ercf}.

The ER formed in the ICF are mainly the Rn isotopes \nuc{212}{Rn}, \nuc{211}{Rn}, \nuc{210}{Rn}, and \nuc{209}{Rn} produced after the $\gamma$-ray emission and after the emission of one, two, and three neutrons from the conjugate nucleus \nuc{212}{Rn}$^*$ [Fig.~\ref{fig:ertot}(b)]. The $\alpha2n$ channel dominates at the lower energies, while at higher energies the $\alpha0n$, $\alpha1n$, and $\alpha2n$ channels give comparable contributions, and the cross section of the $\alpha4n$ channel is small. The large survival probability of \nuc{212}{Rn}$^*$ is related to its relatively small excitation energy and to its large fission barrier, $B_f=18.63$~MeV, in comparison with $B_f=13.01$~MeV for \nuc{216}{Ra}~\cite{Moller2015}. In Fig.~\ref{fig:ertot}(b), the calculated ICF contribution is compared with the difference between the measured cross sections of all ER and of the Ra isotopes. The calculated values are of the same order as this difference: they exceed it by a factor up to about 3 at $\Ecm=116$--134~MeV and agree with it within 35\% at $\Ecm\geq137$~MeV. It should be stressed that the measured difference also includes the products of the emission of protons and $\alpha$ particles from \nuc{216}{Ra}$^*$ formed in the complete fusion. In order to estimate their contribution, the survival probability of \nuc{216}{Ra}$^*$ has been calculated taking into account the emission of neutrons, protons, and $\alpha$ particles. The calculated cross sections of the Fr isotopes formed in the $pxn$ channels do not exceed 0.4~mb, the ones of the Rn isotopes formed in the $\alpha xn$ channels do not exceed about 0.1~mb, and the cross sections of the Ra isotopes change by less than 10\%. Therefore, the Rn isotopes are formed in the \nuc{30}{Si}+\nuc{186}{W} reaction mainly in the ICF.

\begin{figure}[!tb]
\includegraphics[width=\columnwidth]{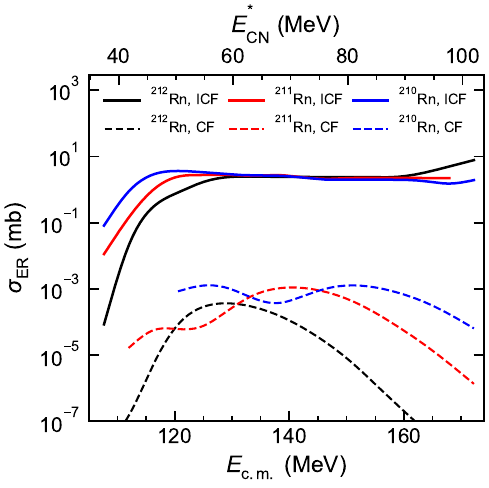}
\caption{Cross sections of the \nuc{212}{Rn}, \nuc{211}{Rn}, and \nuc{210}{Rn} isotopes formed in the \nuc{30}{Si}+\nuc{186}{W} reaction in the ICF (solid curves) and in the de-excitation of \nuc{216}{Ra}$^*$ formed in the complete fusion (dashed curves). The survival probabilities are calculated taking into account the emission of neutrons, protons, and $\alpha$ particles.}
\label{fig:rn}
\end{figure}

This conclusion is especially clear for the heavy isotopes \nuc{212}{Rn} and \nuc{211}{Rn} (Fig.~\ref{fig:rn}). In the de-excitation of \nuc{216}{Ra}$^*$, the isotope \nuc{212}{Rn} can be formed only if the emitted $\alpha$ particle carries away the energy $\varepsilon_\alpha\geq\Ecn+Q_\alpha-S_n(^{212}{\rm Rn})$, where $Q_\alpha=9.53$~MeV is the $\alpha$-decay energy of \nuc{216}{Ra} and $S_n(^{212}{\rm Rn})=7.97$~MeV is the neutron separation energy of \nuc{212}{Rn}~\cite{Moller2016}. For \nuc{211}{Rn}, this energy is lower by $S_n(^{211}{\rm Rn})=7.23$~MeV. At $\Ecm\geq116$~MeV, these conditions require $\varepsilon_\alpha\geq40$~MeV, which is about twice the Coulomb barrier of the $\alpha$+\nuc{212}{Rn} system determining the typical energy of the evaporated $\alpha$ particles. As a result, the calculated cross sections of \nuc{212}{Rn} and \nuc{211}{Rn} formed in the complete fusion do not exceed about $10^{-3}$~mb. In the ICF, these isotopes are formed from the cold conjugate nucleus \nuc{212}{Rn}$^*$: \nuc{212}{Rn} is formed in the partial waves where $E^*_{\rm ICF}<S_n(^{212}{\rm Rn})$, i.e., near the upper boundary of the ICF window [Fig.~\ref{fig:partial}(c)], and \nuc{211}{Rn} is formed after the emission of one neutron. Their cross sections are about 1--3~mb at $\Ecm=120$--160~MeV, i.e., more than three orders of magnitude larger than the ones formed in the complete fusion. The cross section of \nuc{212}{Rn} is included in the sum (\ref{eq:stot}) presented in Table~\ref{tab:res} and Fig.~\ref{fig:ertot}, where it gives 0.6--2.5~mb at these energies. Thus, the observation of the \nuc{212}{Rn} and \nuc{211}{Rn} isotopes in the \nuc{30}{Si}+\nuc{186}{W} reaction can be considered as a signature of the ICF.

\section{CONCLUSIONS}\label{sec:concl}

The formation of the evaporation residues in the \nuc{30}{Si}+\nuc{186}{W} reaction leading to \nuc{216}{Ra}$^*$ has been studied in the framework of the DNS model taking into account the complete fusion and ICF, which is considered as the emission of the $\alpha$ particle from the very mass-asymmetric configuration of the DNS. The capture, complete fusion, quasifission, and ICF cross sections have been calculated by averaging over the orientation angles of both deformed colliding nuclei, and the survival probabilities of the excited \nuc{216}{Ra} and \nuc{212}{Rn} nuclei have been calculated using the statistical model.

The fusion probability calculated for the partial waves contributing to the ER formation, $\Pcn(\ell\leq30\hbar)\approx0.75$, is larger than the value $0.57\pm0.08$ extracted from the experimental data at the same collision energies~\cite{Berriman2001}. This result supports the conclusion that complete fusion is hindered by quasifission in the \nuc{30}{Si}+\nuc{186}{W} reaction with $Z_1Z_2=1036$.

The part of the ICF cross section leading to the ER is 1--2\% of the capture cross section. These ICF events are formed mainly at $\ell=20$--$60\hbar$, and they lead to the relatively cold conjugate nuclei \nuc{212}{Rn}$^*$, which have a large fission barrier. As a result, in spite of the small ICF cross section, the Rn isotopes formed in the ICF give more than half of the calculated ER cross section above $\Ecm\approx128$~MeV, and they account for the slow decrease of the measured total ER cross section in comparison with the one of the Ra isotopes. The calculation of the de-excitation of \nuc{216}{Ra}$^*$ with the emission of protons and $\alpha$ particles shows that the contribution of the complete fusion to the Rn isotopes is small. In particular, the cross sections of the isotopes \nuc{212}{Rn} and \nuc{211}{Rn} formed in the de-excitation of \nuc{216}{Ra}$^*$ do not exceed about $10^{-3}$~mb at these collision energies, and their cross sections in the ICF are larger by more than three orders of magnitude. Therefore, the measurement of the yields of the \nuc{212}{Rn} and \nuc{211}{Rn} isotopes in the \nuc{30}{Si}+\nuc{186}{W} reaction at $\Ecm>120$~MeV, together with the $\alpha$ particles emitted in coincidence, may serve as a test of the DNS mechanism of the ICF.

\begin{acknowledgments}
During the preparation of this work the authors used Claude (Anthropic; Claude Opus 5.5 and Claude Fable 5.1, via Claude Code) to analyze the results of the calculations, to write and test the Python codes used for the data processing and the figures, to check the bibliography, to perform an internal review of the manuscript, and to draft and edit the text under the authors' direction. All numerical results and statements were reviewed and verified by the authors, who take full responsibility for the content of the manuscript.
\end{acknowledgments}

\section*{DATA AVAILABILITY}
The calculated cross sections that support the findings of this article are presented in Table~\ref{tab:res}. Other data are not publicly available and are available from the authors upon reasonable request.

\bibliography{references}

\end{document}